\documentstyle[epsf]{l-aa}   % in [] one has to write 'referee' for 1
\newcommand{\HI}{H$\,${\sc i}}
\newcommand{\HII}{H$\,${\sc ii}}

\newcommand{\kms}{\mbox{km s$^{-1}$}}

\newcommand{\ebv}{\mbox{$E_{B-V}$}}

\newcommand{\msun}{\mbox{${\cal M}_{\odot}$}}
\newcommand{\nor}{\mbox{$n_{\rm orig}$}}
\newcommand{\Jahr}{\mbox{yr}}       % or a, later for PhD ??
\newcommand{\vtabspace}{\noalign{\smallskip}}
\newcommand{\araa}{ARA\&A}   % Annual Review of Astronomy and Astrophys.
\newcommand{\aj}{AJ}         % Astronomical Journal
\newcommand{\aaa}{A\&A}      % Astronomy and Astrophysics
\newcommand{\aas}{A\&AS}     % Astronomy and Astrophys. Supplement Series
\newcommand{\aar}{A\&AR}     % Astronomy and Astrophysics Review
\newcommand{\apj}{ApJ}       % Astronphysical Journal
\newcommand{\apjs}{ApJS}     % Astronphysical Journal Supplement Series
\newcommand{\mnras}{MNRAS}   % Monthly Notices of the Roy. Astron. Society
\newcommand{\pasp}{PASP}     % Publ. of the Astron. Society of the Pacific
\newcommand{\fcph}{Fund. Cos. Phys.}
\newcommand{\ctcn}{CTIO, Contribution No.}
\newcommand{\pnas}{Proc. Nat. Acad. Sci.}
\newcommand{\esome}{ESO Mes\-sen\-ger}
\newcommand{\mmras}{Mem. R. Astron. Soc.}
\newcommand{\dith}{diploma thesis}
\newcommand{\dsth}{Ph.D. thesis}  % Doctor of Sciences thesis
\begin{document}

% ***********************************************************************
% ********************  Title, Authors, Abstract ************************
% ***********************************************************************

\thesaurus{05         % A&A Section 5: Stellar Clusters and Associations
           (08.05.1;  % Stars: early-type,
            08.08.1;  % (Stars:) Hertzsprung Russell (HR) diagram,
            08.12.3;  % Stars: luminosity function, mass function,
            09.02.1;  % ISM: bubbles
            09.09.1 LMC$\,$4;
                      % ISM: individual objects: ...
            11.13.1)  % (Galaxies:) Magellanic Clouds,
            }         % maximal 6 moegliche Angabe
\title{No stellar age gradient inside supergiant shell LMC$\,$4\thanks{Based on
       observations collected at the European Southern Observatory (ESO),
       La Silla, Chile.}
       }   % Ende von "\title"

\author{Jochen M. Braun\inst{1}
        \and
        Dominik J. Bomans\inst{2,1,}\thanks{Feodor Lynen-Fellow of
                                     the Alexander von Humboldt-Foundation}
        \and 
        Jean-Marie Will\inst{1,}\thanks{Present address: Hewlett-Packard GmbH,
                                Herrenberger Str. 130, D-71034 B\"oblingen}
        \and
	    Klaas S. de Boer\inst{1}}

\institute{Sternwarte der Universit\"at Bonn, Auf dem H\"ugel 71,
           D--53121 Bonn, Federal Republic of Germany \and
           University of Illinois at Urbana-Champaign, Department of Astronomy,
           1002 West Green Street, Urbana, IL 61801, USA}

\offprints{J.M. Braun,
           E-Mail: {\tt 'jbraun@astro. uni-bonn.de'}
          }

\date{Received 2nd May, 1997 / accepted August, 1997}

\maketitle

\markboth{J.M. Braun et al.: No stellar age gradient inside supergiant shell
          LMC$\,$4}{}

\begin{abstract}
The youngest stellar populations of a 'J'-shaped region inside the
supergiant shell (SGS) LMC$\,$4 have been analysed with CCD photometry 
in $B$, $V$ passbands.
This region consists of 2 coherent strips, one from the east to west
reaching about $400\;$pc across the OB superassociation LH$\,77$ and
another extending about $850\;$pc from south to north.

The standard photometric methods yield 25 colour-magnitude diagrams (CMDs)
which were used for age determination of the youngest star population by
isochrone fitting. The resultant ages lie in the range from $9\;$M\Jahr\ to
$16\;$M\Jahr\ without correlation with the distance to the LMC$\,$4 centre.
We therefore conclude that there must have been one triggering event for 
star formation inside this great LMC SGS with a diameter of $1.4\;$kpc.

We construct the luminosity function and the mass function of five regions 
consisting of 5 fields to ensure that projection effects don't mask the 
results.
The slopes lie in the expected range ($\gamma \in [0.22;0.41]$ and
$\Gamma \in [-1.3;-2.4]$ respectively). 
The greatest values of the slope occur in the north, which is caused by the
absence of a young, number-dominating star population.

We have calculated the rate with which supernovae (SNe) have exploded in
LMC$\,$4, based on the finding that all stars are essentially coeval. 
A total of 5--7$\cdot10^3$ supernovae has dumped the energy of
$10^{54.5}\;$erg over the past 10$\;$Myr into LMC$\,$4, 
in fact enough to tear the original star-forming cloud apart in the time span
between 5 and 8$\;$Myr after the starformation burst.
We conclude that LMC$\,$4 can have been formed without a contribution from
stochastic self-propagating star formation (SSPSF), although the ring of young
associations and \HII\ regions around the edge have been triggered by the
events inside LMC$\,$4. 

\end{abstract}

% ***********************************************************************
% **************************  Introduction  *****************************
% ***********************************************************************

\section{Introduction}

% *******  Supergiant shells  **********
In the Magellanic Clouds (MCs) giant loops of \HII\ regions
can be recognized in deep pictures taken in H$\alpha$ light.
These \HII\ structures were divided by Goudis \& Meaburn (1978)
into two distinct groups:
shell structures called giant shells (GSs) with diameters of 20--$260\;$pc
and the huge supergiant shells (SGSs) with diameters of 600--$1\,400\;$pc.
Employing unsharp masking techniques and high contrast copying
of the long exposed H$\alpha$ images of Davies et al. (1976, hereafter DEM),
Meaburn and collaborators (see Meaburn 1980) subsequently identified 85 GSs
and 9 SGSs in the Large Magellanic Cloud (LMC) and 1 SGS in the Small
Magellanic Cloud (SMC).

Unlike GSs whose stucture can be explained by the combined activity of
supernova explosions, stellar winds and radiation pressure of the central 
star association(s), SGSs (i.e. structures about $1\;$kpc in diameter) need 
very effective, gigantic mechanisms for their creation.
Such large scale features, which rival in size only with spiral structures,
might be created, according to the appraisal of mechanisms by 
Tenorio-Tagle \& Bodenheimer (1988), by the collision of high velocity clouds 
(HVCs) with the disk of the galaxy or by stochastic self-propagating star 
formation (SSPSF).

The infall of an HVC would cause a well defined velocity deviating from 
the velocity of the original disk gas.
Also, because of the short time scale of such a collision, an almost 
identical age is expected for all stars whose formation was triggered 
by that event.

In the case of SSPSF (see Feitzinger et al. 1981) star formation propagates
from one point (e.g. the centre of a SGS) in all directions (e.g. towards 
the rim of a SGS).
Thus one should see in projection two velocity components in the \HI\ layer
(a receeding and an approaching component) as well as a clear age gradient
of the star populations inside a SGS.

However, the observation of a central area of a SGS with little gas
(i.e. a 'hole' in the \HI\ layer) and of an approximately  $200\;$pc thick
shell of neutral material, which is ionized at the inner edge
(visible as H$\alpha$ filaments) by the radiation from the associations
containing numerous early type stars (see Lucke \& Hodge 1970; Lucke 1974)
in the centre of the SGS, would be consistent with both scenarios.

The importance of understanding the formation and the structure of SGSs
is evident if one realizes how big they are.
They may dominate large portions of a galaxy, in particular of the
smaller irregular galaxies, such as the Magellanic Clouds are.
Knowledge about the creation of SGSs leads to a better knowledge of the
more recent development of the MCs and of their youngest star populations.

% *****  Dynamical models for the SGS LMC$\,$4  *****
The shape and size of LMC$\,$4 is such, that SSPSF has been considered
as the most likely explanation for the formation of this SGS (see Dopita
et al. 1985).
However, observations of star groups along the edge of LMC$\,$4 showed
that the ages of these groups were of the same order as the time needed 
to create the SGS in the case of SSPSF (see Vallenari et al. 1993; 
Petr et al. 1994).

Reid et al. (1987) found from $V$, $I$ photometry of Shapley 
Constellation III (the southern half of the SGS LMC$\,$4, McKibben 
Nail \& Shapley 1953) no clear age gradient, also inconsistent with
a global SSPSF model.

Furthermore, Domg{\"o}rgen et al. (1995) presented an investigation of
LMC$\,$4 based on \HI\ data and IUE spectra.
One result is that there is only one distinct velocity component towards
us with $10\;\kms$ and just a diffuse rear component.
This is not consistent with an undisturbed expanding shell, but it indicates
a break-out of the SGS at the back side of the LMC.

So LMC$\,$4 resembles a cylinder rather than a sphere which should be expected
in a galaxy with an \HI\ scale height well below $500\;$pc.
We note, however, that it is notoriously difficult to derive depth structure 
in a reliable way.

% ***********************************************************************
% ********************  f_lmc4do.tex of jolh77.tex  *********************
% ***********************************************************************
\begin{figure}
\epsfxsize=8.95cm
\centerline{\epsffile{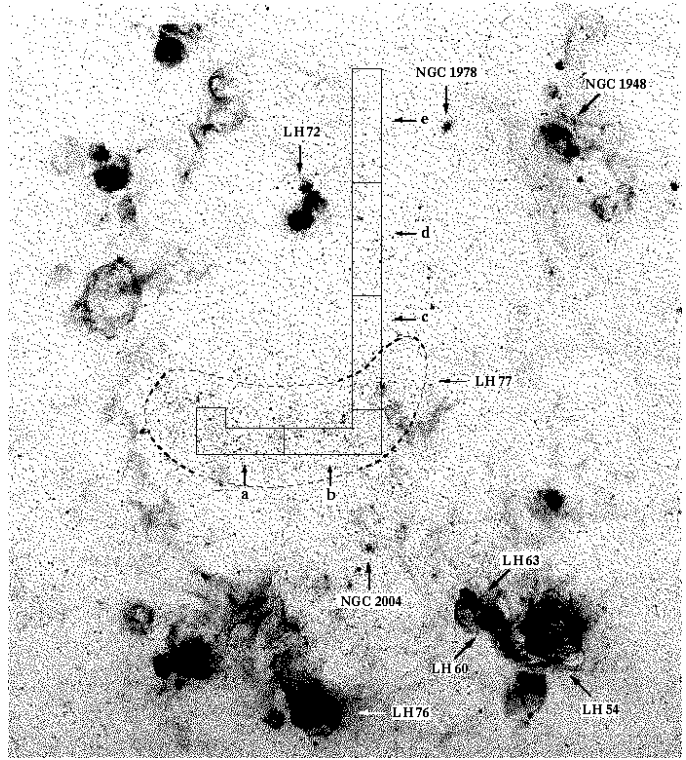}}
\caption[]{H$\alpha$ image of supergiant shell LMC$\,$4 made from a scan
           of a photographic plate taken with the Curtis Schmidt telescope
           at Cerro Tololo (Kennicutt \& Hodge 1986). 
           The locations of the important objects indicated in the text and 
           of the 5 regions of the 1st dataset (named a--e, see Sect. 4) are
           marked}
\end{figure}

All these points show that we still do not well understand the history of
LMC$\,$4.
To improve this situation, photometry of star groups {\it inside} LMC$\,$4 
was taken in 1993 with the goal to derive ages.
This paper presents that data and the results.
Additionally the star formation history is further investigated in Sect. 4
by looking for overlapping age groups in the derived mass functions.

% ***********************************************************************
% ***************  Observations and data reduction  *********************
% ***********************************************************************

\section{Observations and data reduction}

Our data\footnote{
  The entire Tables 1 and 3 of this publication are only available
  electronically, at the CDS (see Editorial in \aaa\ 280, E1,
  1993) or at the Astronomical Institutes of Bonn University
  ({\tt 'ftp ftp.astro.uni-bonn.de'}, see URL
  {\tt 'http://www. astro.uni-bonn.de/{$\,\,\tilde{}\,\,$}jbraun/diploma.html'}
  in the WWW for further information).}
were taken in April, 1993 (during 13 nights in Dutch time) with the $0.91\;$m
Dutch telescope (1st dataset). 
A second, very limited dataset was taken on the 6th February, 1996 with the
$1.54\;$m Danish telescope.

The Dutch telescope was equipped with a $512^2\;$pix$^2$ Tektronix CCD 
(ESO \#33) with a scale factor of $0.443''$ pix$^{-1}$ ($1\;$pix =
$27\;\mu$m) and a total field of view of $3.8' \times 3.8'$,
the Danish telescope with a $2048^2\;$pix$^2$ Loral/Lesser CCD (W11-4 Chip)
with a scale factor of $0.39''$ pix$^{-1}$ ($1\;$pix = $15\;\mu$m).

% ***********************************************************************
% ********************  t_photbv.tex of jolh77.tex  *********************
% ***********************************************************************
\begin{table*}
\caption[]{Part of the table of photometric results around Sk$\,-$67$\,$198 =
           MACS$\,$J0534$-$670\#003 (Magellanic Catalogue
           of Stars: Tucholke et al. 1996, see Table 3).
           The position of the stars in the (rough) general coordinate
           system $(\tilde{x},\tilde{y})$ and on the long exposed $B$ frames
           $(x,y)$ are given with the calibrated $B$, $V$ magnitudes and total
           errors (statistic and systematic) and the dereddened $M_V$ and
           $(B-V)_0$ values.
           The table with 20$\,$812 stars is available electronically
           (see the footnote to Sect. 2)}
\begin{tabular}{cccccccrccr}
\hline
\vtabspace
Field sequence & $\tilde{x}$ & $\tilde{y}$ & $x$   & $y$   & $V$   & $\Delta V$ & \multicolumn{1}{c}{$B-V$} & $\Delta (B-V)$ & $M_V$ & \multicolumn{1}{c}{$(B-V)_0$} \\
number         & [pix]       & [pix]       & [pix] & [pix] & [mag] & [mag]      & \multicolumn{1}{c}{[mag]} & [mag]          & [mag] & \multicolumn{1}{c}{[mag]} \\
\vtabspace
\hline
\vtabspace
$\vdots$ & $\vdots$ & $\vdots$ & $\vdots$ & $\vdots$ & $\vdots$ & $\vdots$ & \multicolumn{1}{c}{$\vdots$} & $\vdots$ & $\vdots$ & \multicolumn{1}{c}{$\vdots$} \\
  2.0030 & 678.1 & 437.7 & 340.9 & 266.8 & 16.145 & 0.074 & $-$0.153 & 0.116 & $-$2.696 & $-$0.263$\;\;$ \\
  2.0031 & 853.4 & 441.3 & 165.7 & 270.4 & 16.130 & 0.075 &    1.697 & 0.119 & $-$2.711 &    1.587$\;\;$ \\
  2.0032 & 757.4 & 450.6 & 261.6 & 279.7 & 11.866 & 0.073 &    0.000 & 0.116 & $-$6.975 & $-$0.110$\;\;$ \\
  2.0033 & 732.9 & 453.7 & 286.2 & 282.8 & 17.990 & 0.079 &    0.918 & 0.130 & $-$0.852 &    0.808$\;\;$ \\
  2.0035 & 799.3 & 458.3 & 219.7 & 287.4 & 16.559 & 0.074 & $-$0.143 & 0.116 & $-$2.282 & $-$0.253$\;\;$ \\
$\vdots$ & $\vdots$ & $\vdots$ & $\vdots$ & $\vdots$ & $\vdots$ & $\vdots$ & \multicolumn{1}{c}{$\vdots$} & $\vdots$ & $\vdots$ & \multicolumn{1}{c}{$\vdots$} \\
\vtabspace
\hline
\end{tabular}
\end{table*}

The 1st dataset contains 25 CCD fields (0--24, see Fig. 3) with an overlap
of about 80$\;$pix.
They form two coherent strips, 
one $400\;$pc strip through the OB superassociation LH$\,$77 from east
to west with 10 fields (i.e. CCD positions which cover an area of
$14.2\;\Box '$ each) and
an $850\;$pc strip from the south to the north with 15 fields, reaching
the rim of the supergiant shell LMC$\,$4 (see Figs. 1 and 3).
The whole 'J'-shaped area is about 298$\;\Box ' = 0.083\;\Box^\circ$ in
size (without the 16{\%} of overlapping regions).
For each position we have long exposed frames of 10$\;$min in the $B$ and 
5$\;$min in the $V$ passband (ESO \#419 and \#420) reaching down to
approximately $V = 20\;$mag (fields 0--4 corresponding to region a, see
Sect. 4) or to $V = 21\;$mag (fields 5--24 corresponding to regions 
b--e) and short exposed frames of $1\;$min in $B$ and $0.5\;$min in $V$, all
with a seeing of 1.3${''}$--2.6${''}$.
The 2nd dataset contains 3 CCD fields ($^+0$, $^+2$ and $^+24$, see
Fig. 3) at the beginning and the end of the area covered 
by the 1st dataset. 
This set was obtained to get a better calibration of the main dataset, so we
took short exposed frames in the $B$ and $V$ passbands (ESO \#450 and \#451)
reaching down to $V = 19\;$mag with a seeing of 1.8${''}$--2.7${''}$.

The data reduction was carried out with MIDAS and the built-in photometry
package DAOPHOT II of Stetson (1987). Additionally some special steps (e.g.
correction of bad double columns and reading the time information of the
header) was done with IRAF.

To calibrate the frames we used the following standard fields of Landolt
(1992): Rubin$\,$149, PG$\,$0918$+$029 and PG$\,$1633$+$099.
Because of wrong time information in the file headers we were only able
to use the standard fields of night 8 and 9 of the 1st dataset (i.e. fields
11--16) and to calibrate the beginning (i.e. fields 0--3) and the end (i.e.
fields 23--24) of the 'J'-shaped region by the 2nd dataset.
To get a homogeneous dataset (see Table 1) we adjusted the magnitude levels
of the fields 4--10 and 17--22 by their overlapping areas.

% ***********************************************************************
% ********************  t_photer.tex of jolh77.tex  *********************
% ***********************************************************************
\begin{table}
\caption[]{Mean total and statistical errors with standard deviation of 
           stellar $V$ magnitudes and $B-V$ colours.
           The number of stars in the corresponding magnitude range
           is given in parentheses}
\begin{tabular}{cc@{\ }c@{\ }c@{\ }c@{\ }ccll}
\hline
\vtabspace
\multicolumn{7}{c}{Magnitude range} &
  \multicolumn{2}{c}{Mean errors} \\
\multicolumn{7}{c}{[mag]} & \multicolumn{2}{c}{[mag]} \\ \vtabspace
\cline{8-9} \vtabspace
\multicolumn{7}{c}{(Number of stars)} & \multicolumn{1}{c}{$\Delta V$}
                                      & \multicolumn{1}{c}{$\Delta (B-V)$} \\
\vtabspace
\hline
\vtabspace
&        &       & $V$ & $<$ & 18 & & $0.076\,(14)$ & $0.120\,(22)$ \\
& \multicolumn{5}{c}{($3\,270$)}  & & $0.014\,(19)$ & $0.021\,(28)$ \\
\vtabspace
&     18 & $\le$ & $V$ & $<$ & 20 & & $0.090\,(31)$ & $0.144\,(46)$ \\
& \multicolumn{5}{c}{($10\,459$)} & & $0.045\,(39)$ & $0.068\,(55)$ \\
\vtabspace
&     20 & $\le$ & $V$ &     &    & & $0.137\,(63)$ & $0.223\,(100)$ \\
& \multicolumn{5}{c}{($7\,083$)}  & & $0.112\,(70)$ & $0.167\,(101)$ \\
\vtabspace
\hline
\end{tabular}
\end{table}

The calibration of the 1st dataset was done with the following equations:
\begin{equation}
V =
\begin{array}[t]{l}
  V_{\rm n} - 3.210\,(13)\;\mbox{mag} + 0.029\,(7) \cdot \\[0.2cm]
  \left[ (B - V)_{\rm n} - 0.450\,(18)\;\mbox{mag} \right]
\end{array}
\end{equation}
\begin{equation}
(B - V) =
\begin{array}[t]{l}
  \left[ (B - V)_{\rm n} - 0.450\,(18)\;\mbox{mag} \right] /
  0.907\,(9)
\end{array}
\end{equation}
with index n indicating the normalization to exposure time $1\;$s and
airmass $0$ and of the shift from point spread function (PSF) to aperture
magnitudes.
The airmass correction was made by means of the atmospheric extinction
coefficients measured on La Silla by the Geneva group (Burki et al. 1995a, b).

The mean errors of our photometry are given in Table 2.
The total error contains all statistical and systematical errors (including
PSF fit, calibration and PSF to aperture shift as their main part) and is
rather an overestimation (see Fig. 2), while the statistical
error from DAOPHOT is a clear overestimation of the reached accuracy.

% ***********************************************************************
% *********************  t_macs.tex of jolh77.tex  **********************
% ***********************************************************************
\begin{table*}
\caption[]{Cross identification of the stars in our analysed area with
           the MACS (Magellanic Catalogue of Stars: Tucholke et al. 1996),
           which may serve as an astrometric reference grid.
           As an example we give the data of 5 stars around
           Sk$\,-$67$\,$198 = MACS$\,$J0534$-$670\#003.
           The field sequence number gives the CCD field (here 2) and the
           sequence number in the original DAOPHOT table, the $x$ and $y$
           coordinates show the position on the long exposed $B$ frames.
           The entire table is available electronically (see the footnote
           to Sect. 2)}
\begin{tabular}{cr@{ }r@{ }r@{}lr@{ }r@{ }r@{}lccc}
\hline
\vtabspace
MACS  & \multicolumn{4}{c}{$\alpha$}           & \multicolumn{4}{c}{$\delta$} &
Field sequence & $x$   & $y$ \\
name  & [$^{\rm h}$ & $^{\rm m}$ & $^{\rm s}$ &] & [$^\circ$ & $'$ & $''$ &] &
        number & [pix] & [pix] \\
\vtabspace
\hline
\vtabspace
$\vdots$         & \multicolumn{4}{c}{$\vdots$} & \multicolumn{4}{c}{$\vdots$} &
 $\vdots$ & $\vdots$ & $\vdots$ \\
J0533$-$669\#053 & 5 & 33 & 59&.941 & $-$66 & 59 & 56&.69 & 2.0202 & 216.1 & 412
.6 \\
J0534$-$670\#002 & 5 & 34 & 02&.625 & $-$67 & 00 & 32&.02 & 2.0175 & 254.6 & 338
.9 \\
J0534$-$670\#003 & 5 & 34 & 03&.094 & $-$67 & 00 & 59&.24 & 2.0032 & 261.6 & 279
.7 \\
J0534$-$670\#004 & 5 & 34 & 03&.522 & $-$67 & 01 & 49&.54 & 2.0117 & 273.7 & 170
.4 \\
J0534$-$670\#005 & 5 & 34 & 04&.026 & $-$67 & 00 & 10&.46 & 2.0047 & 269.8 & 386
.5 \\
$\vdots$         & \multicolumn{4}{c}{$\vdots$} & \multicolumn{4}{c}{$\vdots$} &
 $\vdots$ & $\vdots$ & $\vdots$ \\
\vtabspace
\hline
\end{tabular}
\end{table*}

% ***********************************************************************
% ********************  f_cmdiso.tex of jolh77.tex  *********************
% ***********************************************************************
\begin{figure*}
\epsfxsize=18.0cm
\centerline{\epsffile{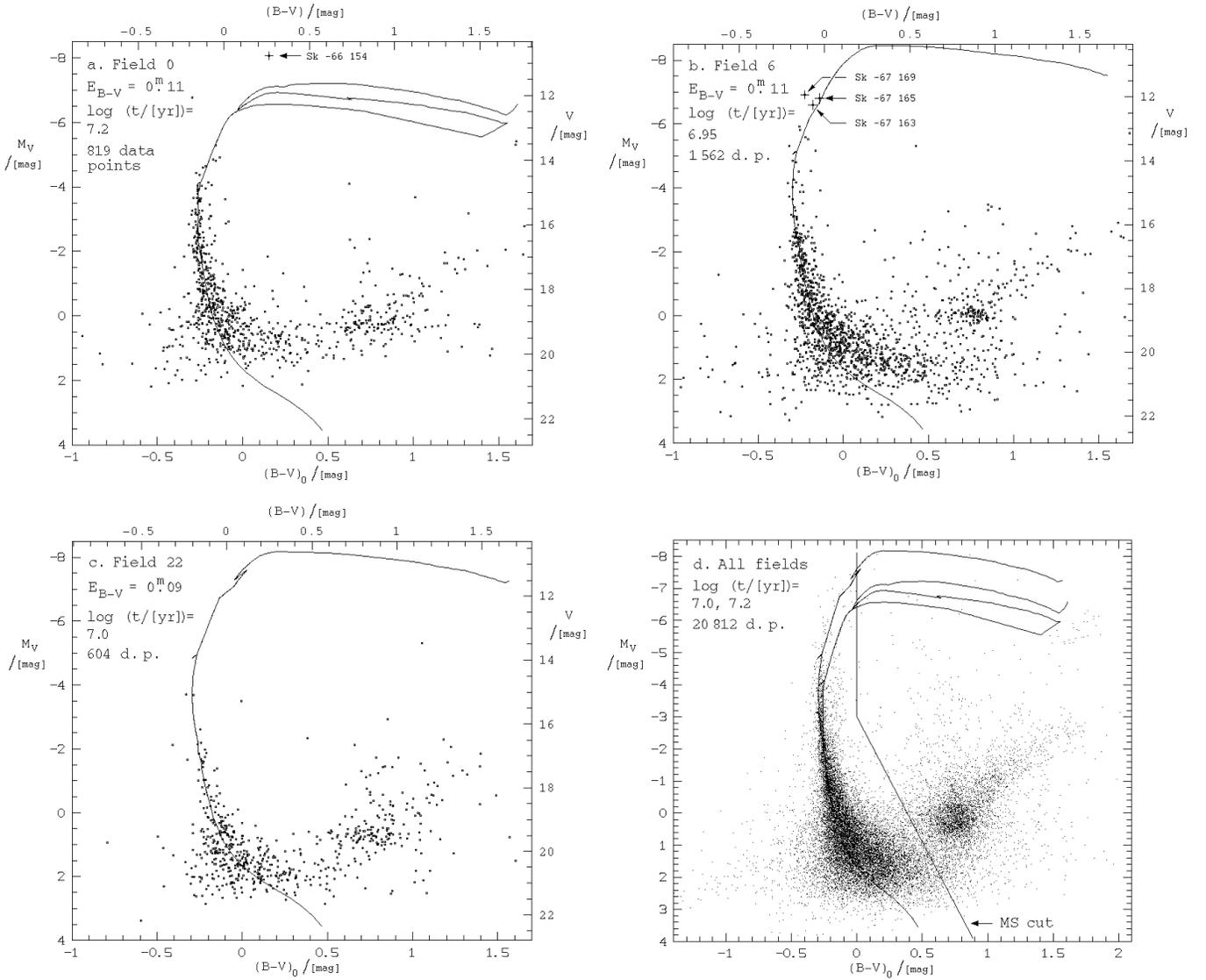}}
\caption[]{CMDs of the fields 0 (a), 6 (b) and 22 (c)
           with marked Sanduleak stars (1969) and of the whole
           analysed area (d) with the rough main sequence cut shown together
           with appropriate Geneva isochrones (Schaerer et al. 1993)}
\end{figure*}

There is no CCD photometry inside LMC$\,$4 in the literature overlapping 
with our data, so we compared the brightest stars of spectral type B0 to A9 
and luminosity class I/Ia marked by Sanduleak (1969) with the $B$, $V$
magnitudes of Rousseau et al. (1978).
%
%
% ***********************************************************************
% *********************  f_l4hwf.tex of jolh77.tex  *********************
% ***********************************************************************
\begin{figure}
\epsfxsize=8.70cm
\epsffile{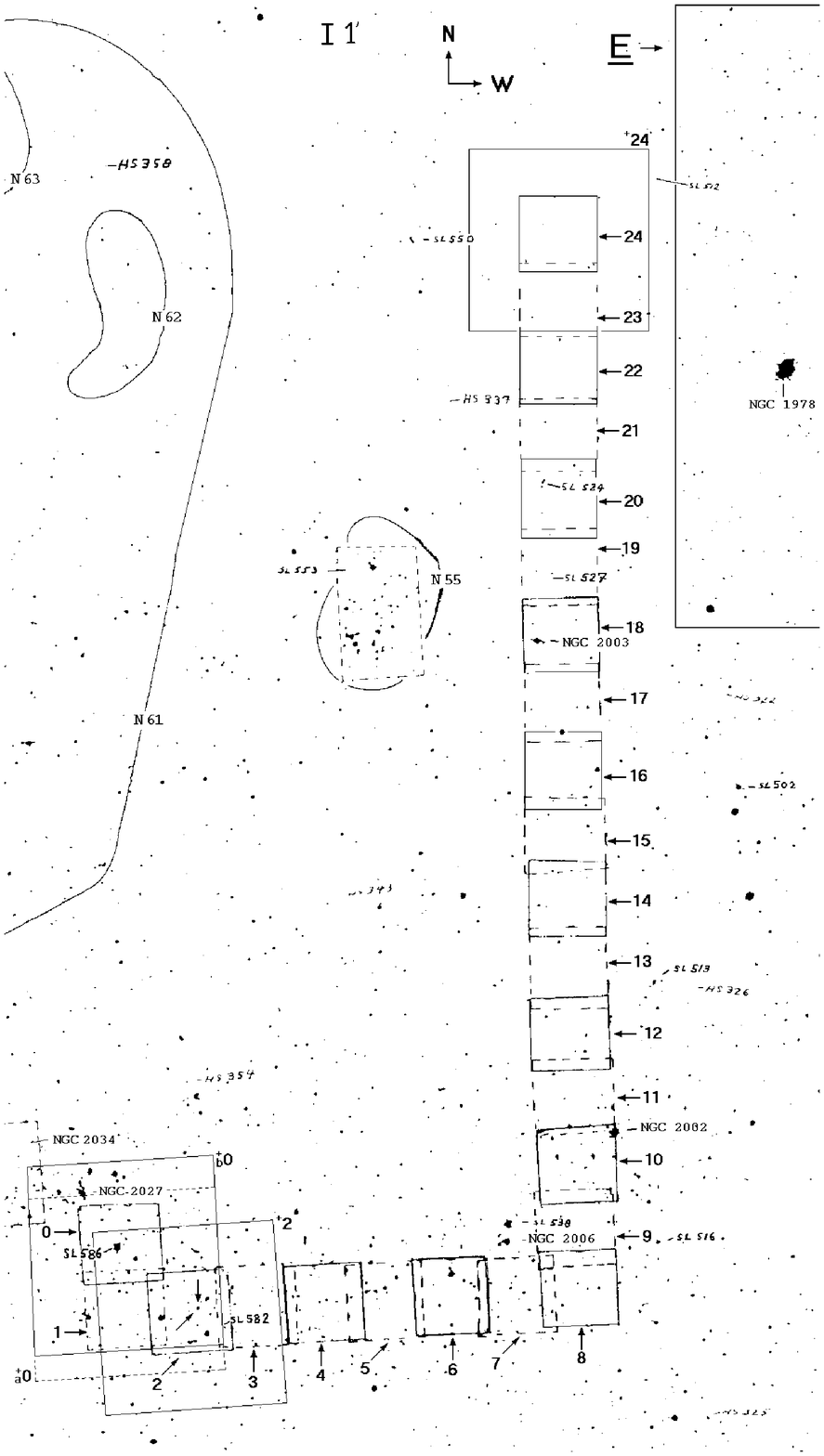}
\caption[]{Mosaic of the central region of LMC$\,$4 out of 4 $V$ charts
           (44, 45, 51, 52) of the LMC atlas of Hodge \& Wright (1967).
           The 25 fields of the 1st dataset, the 3 fields of the 2nd dataset
           and the eastern part of key region E (de Boer et al. 1989, 1991)
           are outlined.
           In field 2 we marked Sk$\,-$67$\,$198 (see Table 3)
           by two arrows}
\end{figure}
This leads to an unexpected high deviation of the 1st dataset, which concerns
only up to five of the brightest (evolved) stars of a CCD field, so it won't
effect the results of the isochrone fit. For all other stars the values of the
1st and 2nd dataset agree within the scope of the errors.

To calculate absolute magnitudes we used the distance modulus of 
$(m-M)_0 = 18.5\;$mag (see Westerlund 1990, and refs. therein) corresponding
to a distance of $50\;$kpc.

For identification purpose we looked for stars having MACS entries
(Magellanic Catalogue of Stars) to get good positions;
%
%
% ***********************************************************************
% ********************  t_fldres.tex of jolh77.tex  *********************
% ***********************************************************************
\begin{table}
\caption[]{Age ($t$), reddening (\ebv) and number of stars ($N_{\ast, BV}$)
           of all 25 CCD fields of the 1st dataset containing $24\,480$ stars
           with $B$, $V$ magnitudes (inclusive multiple counts for stars in the
           overlapping regions)}
\begin{tabular}{rrrrr}
\hline
\vtabspace
Region & Field & \multicolumn{1}{c}{$N_{\ast,BV}$} & \multicolumn{1}{c}{$t$}
  & \multicolumn{1}{c}{\ebv} \\
      & &  & \multicolumn{1}{c}{[M\Jahr]} & \multicolumn{1}{c}{[mag]} \\
\vtabspace
\hline
\vtabspace
 e & 24 &    $313$ &  n &  0.08 \\
   & 23 &    $368$ &  n &  0.08 \\
   & 22 &    $604$ &  n &  0.09 \\
   & 21 &    $819$ &  n &  0.10 \\
   & 20 & $1\,084$ &  n &  0.11 \\ \cline{1-2} \vtabspace
 d & 19 &    $980$ & 11 &  0.11 \\
   & 18 & $1\,132$ & 11 &  0.11 \\
   & 17 & $1\,060$ & 10 &  0.11 \\
   & 16 & $1\,067$ & 10 &  0.09 \\
   & 15 & $1\,120$ & 10 &  0.09 \\ \cline{1-2} \vtabspace
 c & 14 & $1\,166$ & 11 &  0.04 \\
   & 13 &    $698$ & 11 &  0.00 \\
   & 12 &    $904$ & 11 &  0.09 \\
   & 11 &    $998$ & 11 &  0.11 \\
   & 10 & $1\,062$ & 11 & :0.11 \\ \cline{1-2} \vtabspace
 b &  9 & $1\,286$ & 14 & :0.11 \\
   &  8 & $1\,582$ & 13 & :0.11 \\
   &  7 & $1\,746$ & 10 & :0.11 \\
   &  6 & $1\,562$ &  9 & :0.11 \\
   &  5 & $1\,625$ &  9 & :0.11 \\ \cline{1-2} \vtabspace
 a &  4 &    $532$ & 10 & :0.11 \\
   &  3 &    $462$ & 13 &  0.08 \\
   &  2 &    $813$ & 13 &  0.11 \\
   &  1 &    $678$ & 11 &  0.11 \\
   &  0 &    $819$ & 16 &  0.11 \\
\vtabspace
\hline
\vtabspace
\end{tabular}
\hspace*{0.15cm} \begin{minipage}[t]{8.3cm}
':' indicating rough values (see Sect. 3) \\
'n' number of MS stars too small for accurate age determination, but
consistent with $\sim11\;$Myr
\end{minipage} \\
\end{table}
e.g. Sk$\,-$67$\,$198 (Sanduleak 1969) equals
MACS J0534$-$670\#003
(de Boer et al. 1995; Tucholke et al. 1996) with the coordinates (2000):
RA: $5^{\rm h}\,34^{\rm m}\,03.094^{\rm s}$ and
Dec: $-67^\circ\,00'\,59.24{''}$ (see Fig. 3 and Table 3).

Examples of the resulting colour-magnitude diagrams (CMDs) are shown in
Fig. 2.

% ***********************************************************************
% ******************  Determination of the age  *************************
% ***********************************************************************

\section{Determination of the age}

We now proceed and will fit isochrones by eye to the colour-magnitude diagram
of each field separately, with $\log (t/[\Jahr])$ and \ebv\ as fit parameters.
We used the isochrones of the Geneva group (Schaerer et al. 1993) for LMC
metallicity ($Z = 0.008$ or [Fe/H]$=-0.34\;$dex).
These isochrones include the effect of convective core overshooting and are
based on the new opacities.
The use of the isochrones of the Padova group (Alongi et al. 1993) would lead
to slightly younger ages (about $0.05$ in logarithmic age); but we stuck to
the Geneva data to stay consistent with the analyses of other regions in and
around LMC$\,$4 by Olsen et al. (1997), Petr (1994) and Wilcots et al. (1996),
as discussed in Sect. 5 (see Table 6).
In all, we estimate that the accuracy of the determined ages is $\sim 0.1$ in
logarithmic units. 
 
According to Ratnatunga \& Bahcall (1985) the number of foreground stars is 
negligible in respect of the fit. 
%
%
% ***********************************************************************
% ********************  f_lfmfb.tex of jolh77.tex  **********************
% ***********************************************************************
\begin{figure*}
\epsfxsize=18.0cm
\epsffile{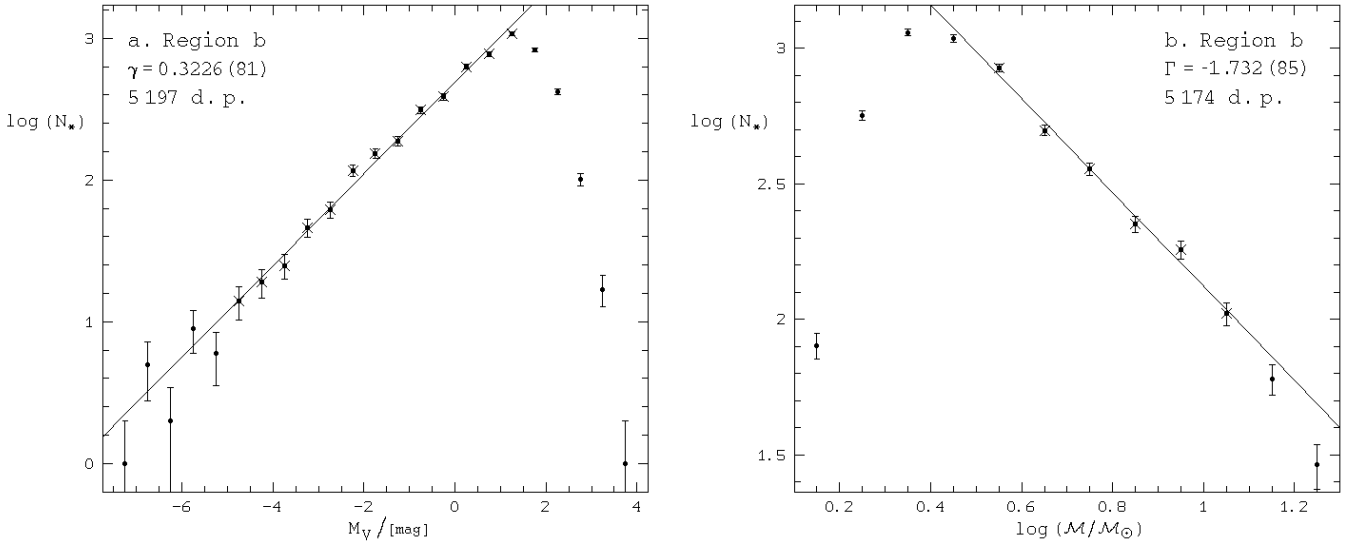}
\caption[]{Luminosity (a) and mass function (b) of region b calculated
           from the Geneva isochrone (Schaerer et al. 1993) of
           $\log (t/[\Jahr]) = 7.0$. The crosses mark the data points
           used in the fit (13 and 6 respectively)}
\end{figure*}
Thus one can expect about 3 stars of our Galaxy towards the LMC in the colour
range of $B-V < 0.8\;$mag and in the apparent visual magnitude range of
$13\;$mag $< V < 19\;$mag (corresponding to the absolute visual magnitude 
range for LMC objects of approximately  $-6\;$mag $< M_V < 0\;$mag).

As examples of this fit to the 1st dataset we show four CMDs in Fig. 2:
the CCD field 0 with the largest age of the analysed area (a),
field 6 with one of the smallest ages (b), and field 22 to demonstrate the
impossibility of fitting an isochrone to the last five fields (i.e. fields
20--24 corresponding to region e, see Sect. 4) because of a poorly
populated upper main sequence (c).
We also show the CMD for the whole 'J'-shaped region (i.e. all CCD fields,
0--24, without overlap) with the appropriate isochrones and the main sequence
cut used to set up the mass function (see Braun 1996 for all CMDs).
We find that all our fields indicate an age between 9 and 16$\;$Myr.  
The results of all 25 fields are listed in Table 4. 

The top of the main sequence region may contain data points for evolved stars.
Given the age derived we expect their number to be small so that they cannot
effect the age determination in a significant way.

The values for the reddening are in accordance with the foreground reddening
of $E_{B-V} \in [0.00;0.15]\;$mag (Oestreicher et al. 1995).
In performing the calibration of the fields in sequence from 3 to 11 we found
small systematic inconsistencies.
Since field 3 and 11 could be calibrated in an absolute manner, we have
interpolated the calibration for the fields in between, leading to less
certain values of the stellar colour and thus of $E_{B-V}$, as given in
Table 4.
We judge the effect of this uncertainty on the ages as negligible.

For all our fields one has to realize that we implicitly assume that all stars
seen in a field belong to the same population. 
However, projection effects can mask age gradients, because in fitting
isochrones one is only sensitive to the youngest star population.
Considering that star formation has been going on
in all regions of the LMC some influence has to be expected.
To handle this problem we investigate the luminosity and mass function for
our region in the next section.

% ***********************************************************************
% ***************  Luminosity function and mass function  ***************
% ***********************************************************************

\section{Luminosity function and mass function}

To get a first hint about the possibility of a superposition of stellar 
populations with different ages - the isochrone fitting would only be 
sensitive for the youngest population if the turn-off points (TOPs) of
the main sequence (MS) are not too different - we combine 5 CCD fields 
to one region (with an area of $59\;\Box '$ on the sky, see Fig. 1). 
For each region we construct the luminosity function (with slope $\gamma$) 
and the mass function (with slope $\Gamma$), as well as for the
whole analysed area.
This was made for equidistant intervals $I$ with the usual approach (see
Scalo 1986):

\begin{equation}
\log \left( N_{\ast,I_{M_V}} \right) = \gamma \cdot
M_V/[\mbox{mag}] + \mbox{const.}
\end{equation}

\begin{equation}
\log \left( N_{\ast,I_{\log({\cal M/M}_\odot)}} \right) = \Gamma \cdot
\log \left( {\cal M}/\msun \right) + \mbox{const.}
\end{equation}

% ***********************************************************************
% ********************  t_regres.tex of jolh77.tex  *********************
% ***********************************************************************
\begin{table}
\caption[]{Per region the number of stars in the photometry is given.
           For the determination of the luminosity function with slope
           $\gamma$ only the indicated number of stars on the main sequence
           is used.
           The same is given for the mass function (with the derived slope
           $\Gamma$).
           No correction for incompleteness at the faint end is made.}
\begin{tabular}{c@{}r@{\quad}rl@{\quad}rl}
\hline
\vtabspace
Region & \multicolumn{1}{c}{$N_{\ast,BV}$} &
\multicolumn{1}{c}{$N_{\ast,{\rm LF}}$} & \multicolumn{1}{c}{$\gamma$} &
\multicolumn{1}{c}{$N_{\ast,{\rm MF}}$} & \multicolumn{1}{c}{$\Gamma$} \\
\vtabspace
\hline
\vtabspace
 e   &  $2\,794$ &  $2\,031$ & $0.406\,(13)$ &  $2\,031$ & $-2.43\,(5)$ \\
 d   &  $4\,606$ &  $3\,579$ & $0.365\,(09)$ &  $3\,573$ & $-2.12\,(4)$ \\
 c   &  $4\,151$ &  $3\,102$ & $0.303\,(11)$ &  $3\,039$ & $-1.65\,(9)$ \\
 b   &  $6\,499$ &  $5\,197$ & $0.323\,(08)$ &  $5\,174$ & $-1.73\,(9)$ \\
 a   &  $2\,762$ &  $2\,129$ & $0.220\,(07)$ &  $2\,118$ & $-1.30\,(8)$ \\ \cline{1-1}
 total$_{\mbox{J}}$ & \quad $20\,812$ & $16\,038$ & $0.318\,(10)$ & $15\,986$ & $-1.72\,(4)$ \\
\vtabspace
\hline
\end{tabular}
\end{table}

As an example we present in Fig. 4 the luminosity and the mass function of
region b.
The translation of brightness to mass for stars on the MS was done with
the best fitting Geneva isochrone of 10$\;$Myr, which can be described with
the numerical relation:
\begin{equation}
{\cal M}/\msun =
\begin{array}[t]{l}
  4.184 - 1.609 \cdot M_V/[{\rm mag}] \\[0.2cm]
  + 0.303 \cdot \left( M_V/[{\rm mag}] \right)^2 \\[0.2cm]
  - 0.036 \cdot \left( M_V/[{\rm mag}] \right)^3 \\[0.2cm]
  - 0.000\,94 \cdot \left( M_V/[{\rm mag}] \right)^4 \\[0.2cm]
  + 0.001\,73 \cdot \left( M_V/[{\rm mag}] \right)^5
\end{array}
\end{equation}
with a rms error of 0.031, beeing valid up to the TOP at
$M_{V,{\rm TOP}} = -4.93\;$mag (corresponding to
${\cal M}_{\rm TOP} = 18.25\;\msun$).

The resulting slopes are, as shown in Table 5, very close to
the standard values of $\gamma \simeq 0.3$ and $\Gamma \simeq -1.35$
(Will 1996).
A  higher negative slope would indicate the superposition of different star
populations (note that incompleteness in the faint bins causes a flatter
distribution than it really would be).

Since the slopes we find are the same as those found in general, we conclude
that there is no significant contribution of older age groups in these fields
up to 300$\;$Myr (equivalent to stars of 3$\;$\msun\ still being on the main
sequence).
Even older star formation events may be present but they would contribute at
fainter levels than the useable limit of our photometry.

% ***********************************************************************
% **********************  The age of LMC 4  *****************************
% ***********************************************************************

\section{The age of LMC$\,$4}

The ages of the various star groups of LMC$\,$4 have been collected
in Table 6.
Note that the location of the various groups can be found in Fig. 1.
For NGC 1948 we list only the newer age determination of Will et al. (1996)
superseding the one by Vallenari et al. (1993).

% ***********************************************************************
% ********************  t_adiop.tex of jolh77.tex  **********************
% ***********************************************************************
\begin{table}
\caption[]{Age ($t$) and reddening (\ebv) of star populations in LMC$\,$4
           sorted from north to south, see Figs. 1 and 3}
\begin{tabular}{lrrl}
\hline
\vtabspace
Object & \multicolumn{1}{c}{$t$}      & \multicolumn{1}{c}{\ebv}  & \multicolumn{1}{c}{Paper}\\
       & \multicolumn{1}{c}{[M\Jahr]} & \multicolumn{1}{c}{[mag]} & \\
\vtabspace
\hline
\vtabspace
 NGC$\,$1978
   &      2$\,$200 &       0.08 & Bomans et al. 1995 \\
 NGC$\,$1948$^a$
   & 5--10 &       0.20 & Will et al. 1996 \\
 LH$\,$72$^b$, north
   &         8--15 & 0.00--0.04 & Olsen et al. 1997 \\
 LH$\,$72$^b$, south
   &             5 & 0.06--0.17 & Olsen et al. 1997 \\
 Region e$^c$
   &             n & 0.08--0.11 & this paper, Table 4 \\
 Region d$^c$
   &        10--11 & 0.09--0.11 & this paper, Table 4 \\
 Region c$^c$
   &            11 & 0.00--:0.11 & this paper, Table 4 \\
 Region b$^d$
   &         9--14 & :0.11 & this paper, Table 4 \\
 Region a$^d$
   &        10--16 & 0.08--:0.11 & this paper, Table 4 \\
 NGC$\,$2004
   &            16 &       0.09 & Sagar \& Richtler 1991 \\
 LH$\,$63$^e$
   &            14 &       0.07 & Petr 1994 \\
 LH$\,$60$^f$
   &             9 &       0.04 & Petr 1994 \\
 LH$\,$54$^g$
   &             6 &       0.10 & Petr 1994 \\
 LH$\,$76$^h$
   &          2--5 &       0.09 & Wilcots et al. 1996 \\
\vtabspace
\hline
\vtabspace
\multicolumn{4}{l}{
\begin{minipage}[t]{8.3cm}
References for acronyms of LMC objects are: \\
LH$\,\ldots\,$ for one of the 122 OB associations and superassociations (Lucke \& Hodge 1970), \\
N$\,\ldots\,$ for one of the 415 emission nebulae (Henize 1956), \\
DEM$\,$L$\,\ldots\,$ for one of the 329 emission nebulae (Davies et al. 1976) \\
and one of the 5 Shapley Constellations (McKibben Nail \& Shapley 1953)
\end{minipage} } \\
\multicolumn{4}{l}{\quad $^a$ NGC$\,$1948 - N$\,$48 - DEM$\,$L$\,$189 - LH$\,$52/53;} \\
\multicolumn{4}{l}{\quad $^b$ LH$\,$72 - N$\,$55 - DEM$\,$L$\,$228;} \\
\multicolumn{4}{l}{\quad $^c$ S--N strip - field 10--24 - region c--e;} \\
\multicolumn{4}{l}{\quad $^d$ E--W strip - field 0--9 - region a--b -
                              LH$\,$77 - main part} \\
\multicolumn{4}{l}{\hspace*{0.48cm} of Shapley Constellation III;} \\
\multicolumn{4}{l}{\quad $^e$ LH$\,$63 - NGC$\,$1974 - N$\,$51$\,$A - DEM$\,$L$\,$201;} \\
\multicolumn{4}{l}{\quad $^f$ LH$\,$60 - NGC$\,$1968 - N$\,$51 - DEM$\,$L$\,$201;} \\
\multicolumn{4}{l}{\quad $^g$ LH$\,$54 - NGC$\,$1955 - N$\,$51$\,$D - DEM$\,$L$\,$192;} \\
\multicolumn{4}{l}{\quad $^h$ LH$\,$76 - NGC$\,$2014 - N$\,$57$\,$A - DEM$\,$L$\,$229.} \\
\end{tabular}
\end{table}

The ages we derived for the 'J'-shaped region are in the range of 9 to
16$\;$Myr.
Also NGC$\,$1948, NGC$\,$2004, LH$\,$72 north, LH$\,$63, and LH$\,$60 are
9 to 16$\;$Myr old.
This seems to suggest that most of LMC$\,$4 was formed as an entity some
9 to 16$\:$Myr ago.
Much younger are LH$\,$54 and LH$\,$76, with an age of 5$\;$Myr,
but these associations are located on the very rim of LMC$\,$4.

N$\,$51, with LH$\,$63, 60 and 54, shows an age gradient (Petr et al. 1994;
Petr 1994), a clear hint at SSPSF on smaller scales.
Much more recent starformation was triggered at the edge of LMC$\,$4,
after the entire interior was in existence for some 5$\;$Myr
(the time between formation of the LMC$\,$4 interior and that of LH$\,$54).
It is interesting to note that 5$\,$Myr is about the time the more massive
stars need to become supernova.

Also LH$\,$72 south is young, but it may have formed indeed more recently,
in the aftermath of events in LH$\,$72 north.

Note that the globular cluster NGC$\,$1978, the rather conspicuous object
near the northern rim of LMC$\,$4 and having an age of approximately $2\;$Gyr
(Bomans et al. 1995), has nothing to do with the creation and/or evolution
of SGS LMC$\,$4.
This statement is valid for all objects older than $70\;$Myr.

Both NGC$\,$1948 and NGC$\,$1978 are part of the $30' \times 30'$ key region
E centered at RA: $5^{\rm h}\,25^{\rm m}$ and Dec: $-66^\circ\,15'$ (de Boer
et al. 1989, 1991; Will et al. 1995).

% ***********************************************************************
% **********************  Summary and Conclusions  **********************
% ***********************************************************************

\section{Toward the history of LMC$\,$4}

LMC$\,$4 contains a substantial population of young stars of age 10$\;$Myr, 
while a quite older ($>300\;$Myr) background population may exist. 
A huge gas cloud must have been present in which the conditions 
were favourable for the burst of star formation $\sim10\;$Myr ago. 
The consequences of that burst are observed today. 
These can be summarized as follows. 
A volume of which we see the projected surface area of about $1\;$kpc$^2$ 
is filled with essentially coeval stars. 
The volume contains little neutral gas in numerous low column density filaments
(Domg{\"o}rgen et al. 1995), and at the same time clearly contains ionzed gas
(see Bomans et al. 1996). 

Inside LMC$\,$4 we have a large number of young main sequence stars. 
From the present day mass function, extended with the same slope to
the mass rich end to also include the original mass rich stars,
we can calculate the number of stars in each mass range 
$\left[ m_{\rm l};m_{\rm u} \right]$ initially present in our 'J'-shaped region:

\begin{equation}
N_{*,{\rm J}} =
\begin{array}[t]{l}
\int\limits_{m_{\rm l}}^{m_{\rm u}} 10^{4.972\,(22)} \cdot ({\cal M}/\msun)^{-2.716\,(44)}\;d{\cal M}
\end{array}
\end{equation}

A fit to the Geneva evolutionary tracks (Schaerer et al. 1993) for an initial
mass ${\cal M} > 12\;\msun$ yields a stellar lifetime relation of:

\begin{equation}
t_{*} =
\begin{array}[t]{l}
(86 \pm 10)\;{\rm Myr} \cdot ({\cal M}/\msun)^{-0.722\,(62)}
\end{array}
\end{equation}

Stars with an initial mass above $8\;\msun$ will become supernovae of Type II. 
Combining Eqs. (6) and (7) with $m_{\rm u} = 125\;\msun$ we get
the number of supernovae after a given time:
\begin{equation}
N_{*,{\rm J,SN}} (t) \simeq
\begin{array}[t]{l}
1.4 \cdot (t/[{\rm Myr}])^{2.38}
\end{array}
\end{equation}

For star populations one would expect all stars with
${\cal M} \in [18.3;125]\;\msun$ to have exploded into SNe in the first
10$\;$Myr.
Thus we get $N_{*,{\rm J}, {\rm SN}} = 320 \pm 30$,
for roughly 5\% of the LMC$\,$4 area.
Extrapolating from the 'J'-shaped region to the entire interior of LMC$\,$4
means that in this SGS of 10$\;$Myr age about 5--7$\cdot 10^3$ supernovae
have exploded. 
With an average supernova energy output of 10$^{51}\;$erg the total 
number of past SNe will have dumped at least 10$^{54.5}\;$erg into LMC$\,$4. 

On the other hand, the supernova rate right after starformation is very small
and Eq. (8) indicates that after 5$\;$Myr only $\sim$ 200 SNe have
exploded inside LMC$\,$4, or less than 10\% of the total in 10$\;$Myr.

At present there is little neutral gas inside LMC$\,$4, 
whereas the structure must have had lots of 
relatively dense neutral gas for the star formation. 
We can estimate the minimum energy input required to dissolve the birth cloud. 
Assuming a thickness of originally 500$\;$pc with a gas density \nor\ in the
volume $V$, the energy needed for total ionization is 
$V \cdot \nor \cdot 13.6\;$eV = $\nor \cdot 10^{53.5}\;$erg cm$^3$. 
A further calculation shows that roughly $\nor \cdot 10^{52}\;$erg cm$^3$ can
accelerate all particles of the entire birth cloud to 100$\;$\kms, enough to
disrupt the cloud. 
This energy is easily provided by the supernovae. 
In fact, a substantial fraction of the energy needed is already released
between 5 and 8$\;$Myr.

Summarizing, based on the recognition that all young stars of LMC$\,$4 are 
nearly coeval at 10$\;$Myr, and on the sequel that supernovae will go off
everywhere inside LMC$\,$4 at a fair rate dumping energy rather evenly in
the birth cloud, we can explain the structure of LMC$\,$4 as we see it today. 

In consequence of that, once the first supernovae occur, their individual
(but soon the collective) blast waves may trigger star formation at the edges.
This fits with the age derived for LH$\,$72 south. 
Starformation at the edges of LMC$\,$4 will have taken place at very recent
times as secondary process.
We expect that an age determination for the stars in any of the \HII\ regions
to the east of LMC$\,$4 will show very young ($<$ 5$\;$Myr) ages, like the ones
at the NW (Will et al. 1996) and the SW (Petr et al. 1994). 

The present supernova rate inside LMC$\,$4 as calculated from our equations is
about 1 per 670$\;$yr which is comparable to the SN rate of the whole LMC
derived from counting supernova remnants (Chu \& Kennicutt 1988).
Our mass function and main sequence lifetime indicate an increase of the SN
rate to 1 per 120$\;$yr in 25$\;$Myr from now, at a time when the stars of
originally 8$\;\msun$ will explode.

% ***********************************************************************
% **********************  Summary and Conclusions  **********************
% ***********************************************************************

\section{Summary and Conclusions}

The ages found for the stars inside LMC$\,$4 lie in the range of $9\;$Myr to
$16\;$Myr without
a visible correlation with the distance to the LMC$\,$4 centre (see
Table 4 and Fig. 3).
This means that there must have been one triggering event for star
formation inside the greatest SGS of the LMC with a diameter of $1.4\;$kpc,
whereas at the rim of LMC$\,$4 there is evidence for SSPSF (Petr et al. 1994;
Petr 1994).
Thus SSPSF can't be the creation mechanism of LMC$\,$4 as claimed by
the first models (see e.g. Dopita et al. 1985),
even if one is able to find examples of SSPSF in this region on scales below
$150\;$pc.

% ***********************************************************************
% *************************  Acknowledgements  **************************
% ***********************************************************************

\acknowledgements
We thank Ger van Rossum for participation in the early stages of this
project and organizing the observations of the 1st dataset.
JMW thanks the ESO staff for their help in obtaining the 2nd dataset.

JMB, JMW and DJB acknowledge support from the Deutsche
Forschungsgemeinschaft (DFG), in the frame of the Graduiertenkolleg
"The Magellanic System and Other Dwarf Galaxies"
(GRK 118/2-96).
DJB is grateful for a Feodor-Lynen-Fellowship of the Alexander von
Humboldt-Foundation.

We thank Antonella Vallenari for discussions, Martin Altmann for critically
reading the manuscript and Rob Kennicutt for the scan used in Fig. 1.

% ***********************************************************************
% ****************************  References  *****************************
% ***********************************************************************

% ***********************************************************************
% *********************  t_bib.tex of jolh77.tex  ***********************
% ***********************************************************************

\end{document}